# Multiferroic Magnon Spin-Torque Based Reconfigurable Logic-In-Memory

Yahong Chai[1,*], Yuhan Liang[1,2,*], Cancheng Xiao[1,*], Yue Wang[2], Bo Li[3], Dingsong Jiang[1], Pratap Pal[4], Yongjian Tang[5], Hetian Chen[2], Yuejie Zhang[1], Witold Skowroński[6], Qinghua Zhang[7], Lin Gu[2], Jing Ma[2], Pu Yu[8], Jianshi Tang[1], Yuan-Hua Lin[2,†], Di Yi[2,†], Daniel C. Ralph[5,9], Chang-Beom Eom[4], Huaqiang Wu[1], Tianxiang Nan[1,†]

1   School of Integrated Circuits and Beijing National Research Center for Information Science and Technology (BNRist), Tsinghua University, Beijing 100084, China

2   School of Materials Science and Engineering, Tsinghua University, Beijing 100084, China

3   Institute for Advanced Study, Tsinghua University, Beijing, 100084, China

4   Department of Materials Science and Engineering, University of Wisconsin-Madison, Madison, Wisconsin 53706, United States

5   Laboratory of Atomic and Solid State Physics, Cornell University, Ithaca, NY 14853, USA

6   Institute of Electronics, AGH University of Science and Technology, Kraków 30-059, Poland

7   Beijing National Laboratory for Condensed Matter Physics, Institute of Physics, Chinese Academy of Sciences, Beijing 100190, China

8   Department of Physics, Tsinghua University, Beijing 100084, China

9   Kavli Institute at Cornell for Nanoscale Science, Ithaca, NY 14853, USA

* These authors contributed equally.

† Email: linyh@tsinghua.edu.cn, diyi@mail.tsinghua.edu.cn, nantianxiang@mail.tsinghua.edu.cn

**Abstract**

**Magnons, bosonic quasiparticles carrying angular momentum, can flow through insulators for information transmission with minimal power dissipation. However, it remains challenging to develop a magnon-based logic due to the lack of efficient electrical manipulation of magnon transport. Here we present a magnon logic-in-memory device in a spin-source/multiferroic/ferromagnet structure, where multiferroic magnon modes can be electrically excited and controlled. In this device, magnon information is encoded to ferromagnetic bits by the magnon-mediated spin torque. We show that the ferroelectric polarization can electrically modulate the magnon spin-torque by controlling the non-collinear antiferromagnetic structure in multiferroic bismuth ferrite thin films with coupled antiferromagnetic and ferroelectric orders. By manipulating the two coupled non-volatile state variables—ferroelectric polarization and magnetization—we further demonstrate reconfigurable logic-in-memory operations in a single device. Our findings highlight the potential of multiferroics for controlling magnon information transport and offer a pathway towards room-temperature voltage-controlled, low-power, scalable magnonics for in-memory computing.**

In-memory computing, utilizing non-volatile memories capable of performing both information storage and logic operations within the same device, holds the promise for empowering artificial intelligence (AI) with significantly reduced energy consumption[1-4]. Existing logic-in-memory devices that have been implemented operate mainly based on charge transport, a process that inevitably gives rise to joule heating. On the other hand, information processing and transmission using magnons[5,6] as information carriers is a promising route for developing spin-based logic and memory devices[6-12] with low-dissipation, since magnons can transport spin in ferrimagnetic and antiferromagnetic insulators without involving moving electrons[13-17]. Incoherent magnons can simply be electrically (and thermally) excited in dc electronic circuits[18-21], making them compatible with current semiconductor technology. For practical applications, the implementation of magnon logic operations using gate voltages is necessary[22-25]. Current technology to manipulate magnon current transport at room temperature mainly relies on magnetic fields that can reorientate the magnetic ordering or modulate the magnetic domain structure[14,26].

An alternative approach involves the utilization of multiferroic materials[27-29] for magnon transport, where the magnetoelectric coupling enables the control of the magnetic order through ferroelectric switching. In the model system of multiferroic BiFeO$_3$[30,31], theoretical predictions suggest the potential for controlling the magnon dispersion by magnetoelectric coupling between the ferroelectric polarization $\vec{P}$ and Néel order $\vec{L}$[32], while experimental results demonstrated electrically tunable spin wave group velocities[33]. Recent studies also show thermally excited magnon currents in BiFeO$_3$ thin films in a longitudinal configuration that are modulated by switching the canted magnetic moment via the ferroelectric polarization[34]. These findings highlight multiferroic materials as an ideal platform for voltage-controlled magnon logic operations. However, integrable magnon-based logic devices with electrically excited incoherent magnons are yet to be realized.

Here, we propose and demonstrate a multiferroic magnon spin-torque (MMST) device for magnon-based reconfigurable logic operations. The device comprises multiple ferromagnetic/multiferroic BiFeO$_3$ memory cells that are positioned on a shared spin-current channel, as shown in Fig. 1a. A charge current pulse ($I_w$) flowing through the channel induces spin accumulation with polarization $\vec{\sigma}$ at the multiferroic bottom interface through the spin-Hall effect or the Rashba-Edelstein effect[35,36], which can excite antiferromagnetic magnon modes depending on the orientation of $\vec{L}$, as the spin injection transmitted to magnons is proportional to $\vec{\sigma} \cdot \vec{L}$[37,38]. As the magnons (carrying the spin-polarized angular momentum from the bottom layer) diffuse across the multiferroic layer to the ferromagnet's bottom interface, they exert a magnon-mediated spin torque (MST) to control the magnetic moment[17,39], enabling a non-volatile writing of spin information to multiple cells on the current channel in parallel. For magnon logic operations, a gate voltage ($V_G$) pulse is applied across the multiferroic BiFeO$_3$, which modulates the antiferromagnetic structure by the strongly coupled $\vec{P}$ and $\vec{L}$ as schematically shown in the pseudo-cubic unit cell of BiFeO$_3$ (Fig. 1b). Bulk BiFeO$_3$ exhibits non-collinear antiferromagnetic order with cycloid structure due to the Dzyaloshinskii-Moriya interaction[33]. For thin films grown on substrates such as DyScO$_3$, the cycloid propagation direction $\vec{k}$ and $\vec{P}$ are orthogonally coupled (Supplementary Fig. 1)[40]. As the excited magnons in BiFeO$_3$ is proportional to $\vec{\sigma} \cdot \vec{L}$ integrated over the cycloid structure (see Supplementary Notes), the modulation of spin cycloid structure in BiFeO$_3$ leads to a non-volatile control of the magnon spin transport.

The device heterostructure is composed of a ferromagnetic multilayer [Pt/Co]$_N$, a multiferroic BiFeO$_3$ layer and a spin-current source SrRuO$_3$ layer which has a large spin Hall conductivity[41,42], as shown in Fig. 1c. We grew epitaxial SrRuO$_3$/BiFeO$_3$ heterostructures on orthorhombic (o) (110)$_o$ DyScO$_3$ substrates (see Methods for details). Piezoelectric force microscopy (PFM) reveals a two-variant stripe ferroelectric domain structure with 71° domain walls (see Supplementary Fig. 2), consistent with previous reports[40,43]. Subsequently, we deposited ferromagnetic multilayer Pt(2)/[Co(0.4)/Pt(0.92)]×3/Co(0.4)/Pt(2) with a robust perpendicular magnetic anisotropy (numbers in parentheses indicate film thickness in nanometer, abbreviated as PtCo) onto BiFeO$_3$ (see Methods for details). The cross-sectional high-angle annular dark-field scanning transmission electron microscope (HAADF-STEM) images of the tri-layer (Fig. 1d) reveals a high crystalline quality and well-defined interfaces between both SrRuO$_3$/BiFeO$_3$ and BiFeO$_3$/PtCo which is crucial for the efficient spin transmission.

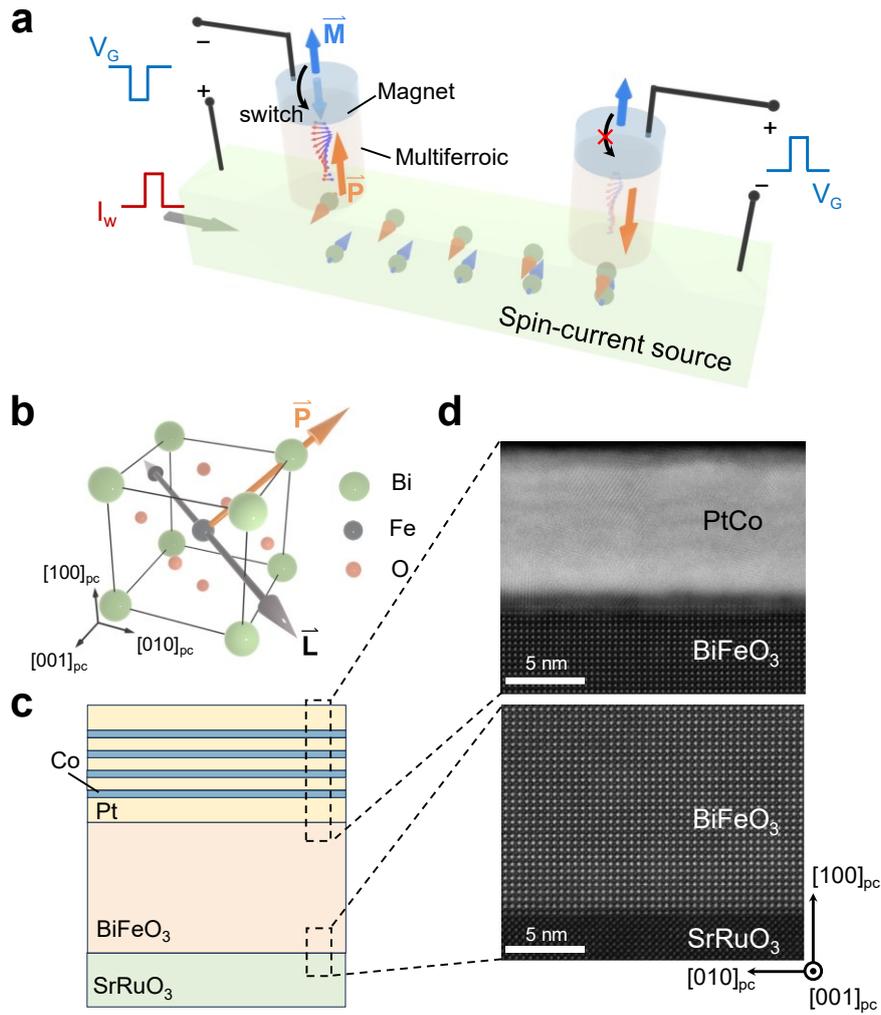

**Figure 1. Working principle and layered structure of the proposed multiferroic magnon spin-torque device. a**, Schematic illustration of the proposed MMST device, in which ferromagnetic/multiferroic junctions are positioned on top of a spin-current source channel. An in-plane charge current pulse $I_w$ generates an out-of-plane magnon current by the spin-Hall effect that induces magnetization (blue arrow) switching through MST. The magnon current can be controlled by the ferroelectric polarization (orange arrow) of the multiferroic layer by gate voltage pulses $V_G$. **b**, Unit cell of $BiFeO_3$ with strongly coupled ferroelectric polarization $\vec{P}$ (orange arrow) and Néel order $\vec{L}$ (gray arrow). **c**, Structure of the $SrRuO_3/BiFeO_3/PtCo$ stack. **d**, HAADF-STEM image of the $SrRuO_3/BiFeO_3/PtCo$ stack, highlighting the $PtCo/BiFeO_3$ interface (upper panel) and the $BiFeO_3/SrRuO_3$ interface (lower panel).

To verify the MMST device concept, we first studied the magnon transport across a multiferroic layer and the MST-induced magnetization switching in a 11 nm SrRuO$_3$/120 nm BiFeO$_3$/PtCo tri-layer with a Hall-bar structure. Fig. 2a illustrates the experimental setup, where the I$_w$ applied along $\vec{x}$ in SrRuO$_3$ generates a spin accumulation at the interface that excites magnon modes in BiFeO$_3$. When reaching the ferromagnetic layer PtCo, these magnons exert a field-like MST ($\vec{\tau}_{m,FL} \propto \vec{M} \times \vec{\sigma}$) and damping-like MST ($\vec{\tau}_{m,DL} \propto \vec{M} \times \vec{\sigma} \times \vec{M}$) on the magnetization $\vec{M}$, where $\vec{\sigma}$ is the spin polarization along $\vec{y}$[17]. Fig. 2b shows the optical micrograph of the device and the measurement configuration. The measured anomalous Hall resistance loop (as a function of external magnetic field along $\vec{z}$, H$_z$) is depicted in Fig. 2c, confirming a perpendicular magnetic anisotropy of PtCo. Figure 2d shows the current pulse-induced switching of perpendicular magnetization of PtCo with an in-plane magnetic field H$_x$=10 mT for a deterministic magnetization switching[44]. Reversing H$_x$ results in the reversal of the MST-induced switching polarity, consistent with the symmetry of the damping-like spin-torque[45,46], ruling out possibilities of magnetization switching induced by Joule heating or Oersted field. We further excluded the self-switching of magnetization in PtCo due to the compositional gradient[47], as no current-induced switching is observed in a control sample of bare PtCo (see Supplementary Fig. 3). Over 90% magnetization could be switched by MST, demonstrating an efficient magnon transport through multiferroic BiFeO$_3$. The threshold switching current density is about 3×10$^6$ A/cm$^2$, comparable to that of the heavy metal/ferromagnetic metal system[48]. The linear dependence[49] of the threshold switching current (I$_c$) on H$_x$ plotted in Fig. 2e (see Supplementary Fig. 3 for the switching hysteresis with different H$_x$) confirms again that the observed magnetization switching is governed by the MST.

The MST efficiency $\xi_{m,DL}$ was measured using other independent spin-torque measurements on a 11 nm SrRuO$_3$/120 nm BiFeO$_3$/5 nm NiFe sample (see Methods for details). Through spin-torque ferromagnetic resonance (ST-FMR)[44] (see Supplementary Fig. 4) and second harmonic Hall voltage (SHHV) measurements[50] (see Supplementary Fig. 5), we estimated the $\xi_{m,DL}$ ranging from 0.012 to 0.027. This $\xi_{m,DL}$ is comparable to the spin-torque efficiency in our SrRuO$_3$ and those reported by others[41,51], which demonstrates a strong MST in the tri-layers with multiferroic BiFeO$_3$ exceeding 100 nm.

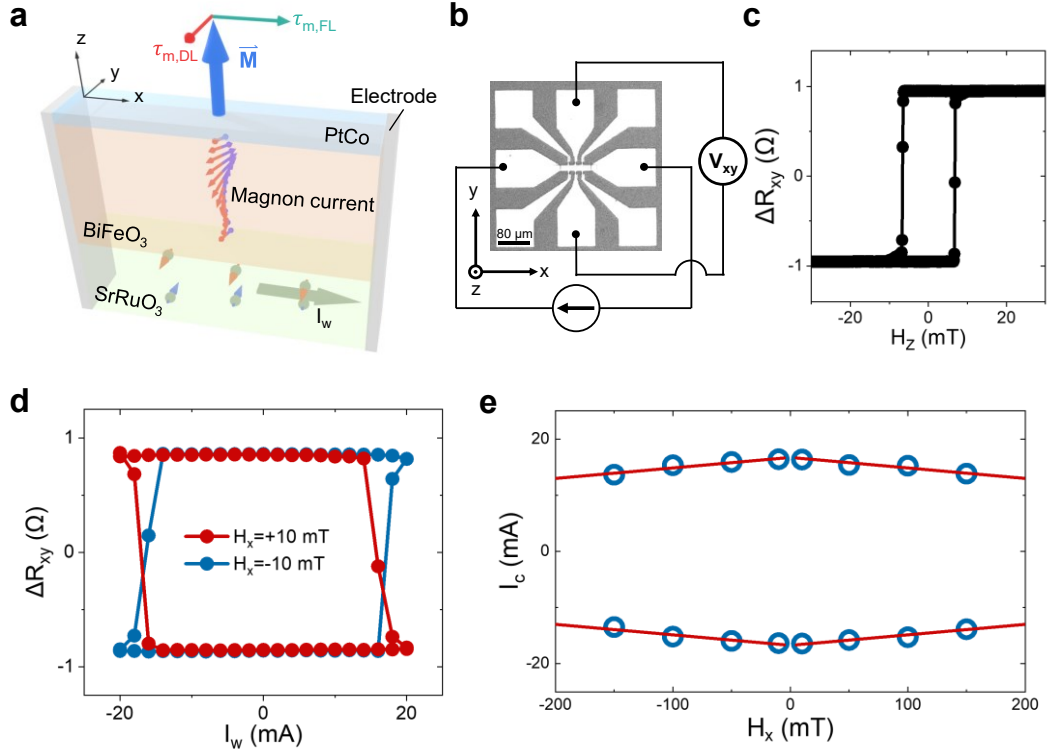

**Figure 2. MST-induced switching of perpendicular magnetization at room temperature. a**, Schematic diagram of the MST-induced perpendicular magnetization switching. The electrodes make contact to both the SrRuO$_3$ and the PtCo layers, so that the applied current flows through both layers in parallel. Red and green arrows represent the damping-like and the field-like components of MST, respectively. The blue arrow denotes the magnetization of the PtCo. **b**, Optical micrograph of a Hall-bar device and the measurement setup for MST-induced switching. **c**, Hall resistance (R$_{xy}$) for the device of 11 nm SrRuO$_3$/120 nm BiFeO$_3$/PtCo as a function of out-of-plane magnetic field H$_z$. **d**, MST induced switching in the device with the presence of an in-plane magnetic field H$_x$ = ±10 mT. **e**, Threshold current for magnetization switching I$_c$ as a function of H$_x$. The blue circles represent experimental data, and the red lines show the linear fitting.

Having established the MST-induced magnetization switching, we proceed to investigate the in-situ voltage control of magnon transport in the proposed MMST device (Fig. 1a). To construct this device, we patterned the SrRuO$_3$ layers into a shared spin-current channel, onto which we etched BiFeO$_3$/PtCo layers into multiple circular micro-pillars as individual bit-cells. As illustrated in Fig. 3a, a global I$_w$ applied to the channel can switch the magnetization of PtCo in the cells through MST, while a local voltage pulse $V_G$ applied to the selected cell across the BiFeO$_3$ layer can reverse the ferroelectric polarization. To probe the magnetization of PtCo, we employed polar magneto-optical effect (MOKE) microscopy, which provides contrast to the out-of-plane component of the magnetization ($M_z$). An optical image of the fabricated device is shown in Fig. 3b, where the probe tip on the pillar and the SrRuO$_3$ layer serve as the top and bottom electrodes for applying $V_G$. Figure 3c exhibits a well-defined hysteresis loop of the PFM signal for a BiFeO$_3$/PtCo cell as the out-of-plane voltage is swept, demonstrating the presence of robust ferroelectricity and two distinct ferroelectric polarization states for BiFeO$_3$ in the MMST device structure.

In Fig. 3d and Fig. 3e, we provide evidence of the voltage-controlled MST. MOKE images were captured in the device consisting of three memory cells with different applied current pulse intervals and two distinct ferroelectric polarization states. The magnetization of the cells was initially set to +$M_z$ (bright MOKE contrast) or -$M_z$ (dark MOKE contrast) by applying out-of-plane magnetic field pulses. When the ferroelectric polarization of all three cells was initially set downwards, we found that a minimum $I_w$ of 8 mA can simultaneously switch the magnetization from +$M_z$ to -$M_z$ (or -$M_z$ to +$M_z$), depending on the polarity of $H_x$ (Fig. 3d). This behavior aligns with the MST-induced magnetization switching discussed in Fig. 2d. To control the MST in a selected cell, $V_G$=-10 V is applied to reverse the out-of-plane component of ferroelectric polarization ($P_z$) of the middle cell from downwards to upwards, while the polarization of other two cells (on the sides) remain downwards. In contrast to Fig. 3d, we observed that a reduced $I_w$ of 7 mA is capable of partially switching (more than 50% of the total area) the magnetization of the middle cell, while the magnetization of the other two cells remains unswitched (Fig. 3e). This indicates a modulation of $I_c$ for MST-induced switching by the ferroelectric polarization, which is estimated to be approximately 14% (ratio = $\frac{I_c^\downarrow - I_c^\uparrow}{I_c^\uparrow} \times 100\%$, where $I_c^\uparrow$ and $I_c^\downarrow$ represents the $I_c$ as ferroelectric polarization is upwards or downwards, respectively). By increasing $I_w$ to 8 mA, the magnetization of all cells is switched, regardless of their ferroelectric polarization direction.

The modulation of $I_c$ does not dependent on the magnetization switching polarity, as shown in Fig. 3e (upper panel: from +$M_z$ to -$M_z$, lower panel: from -$M_z$ to +$M_z$). This observation excludes possible extrinsic effects such as magnetic domain wall pinning in the cell, which could affect the $I_c$ differently depending on the switching polarity. Additionally, we ruled out the voltage-controlled magnetic anisotropy effect[52], as we observed negligible variation in the magnetic hysteresis loop of the cell before and after applying $V_G$ (see Supplementary Fig. 6). A possible mechanism of the voltage-controlled MST in the BiFeO$_3$ film with two-variant domain structure could involve the change of ferroelectric domain structure driven by $V_G$[43] (see Supplementary Fig. 2 for evolution of PFM under voltages). Our model reveals that the magnon transport is quite different in the two domains separated by 71° domain walls, which have the cycloid propagation direction $\vec{k}$ orthogonal to each other (see Supplementary Notes and Supplementary Fig. 1). Therefore, the domain structure provides for an additional handle to modulate the MST.

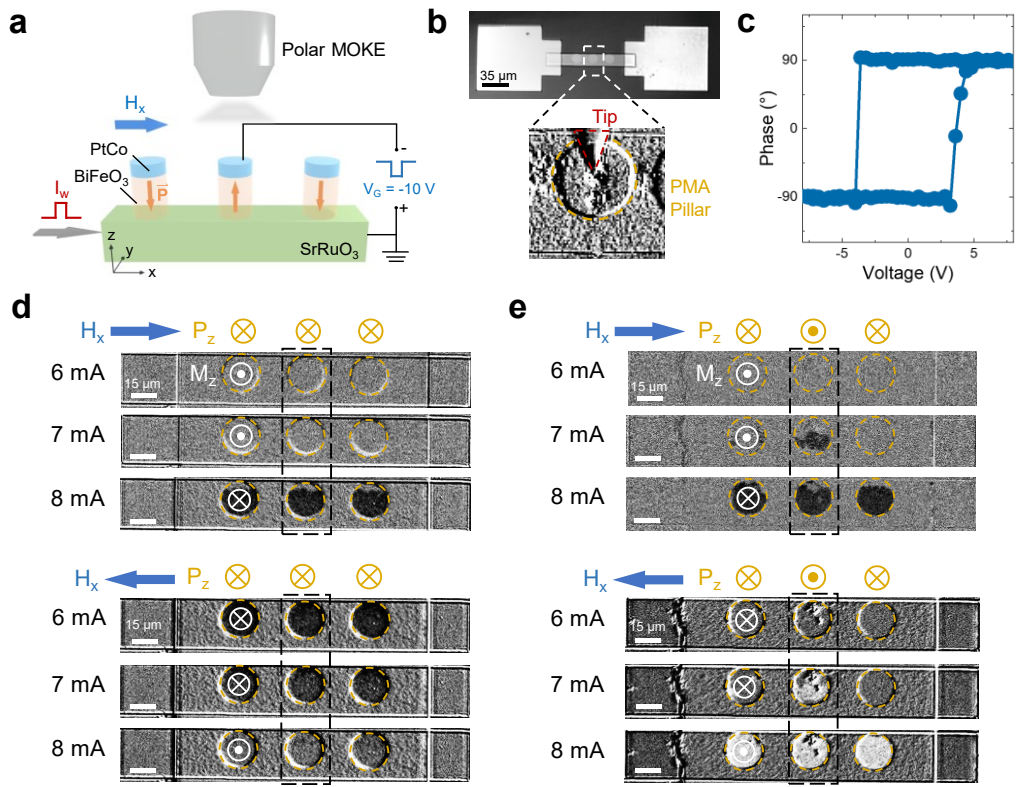

**Figure 3. In-situ voltage control of magnon spin-torque. a**, Schematic for the in-situ voltage control of MST-induced switching in a device comprising 100 nm thick BiFeO$_3$/PtCo memory cells on a 11 nm SrRuO$_3$ spin-current channel. Polar MOKE microscopy is used to probe the $M_z$ of the cell which gives bright (+$M_z$) and dark (-$M_z$) contrasts. The ferroelectric polarization (orange arrow) of cells can be switched by applying $V_G$. **b**, Optical micrograph of the fabricated device. The lower panel shows a zoom-in view of the cell and a probe tip employed for the application of $V_G$. **c**, Out-of-plane phase signal of PFM as a function of the applied voltage. **d** and **e**, MOKE images

illustrating voltage-controlled MST-induced switching in three cells. The ferroelectric polarization of the middle cell circled by the black dashed box is downwards (**d**) or upwards (**e**) by applying $V_G$ before the injection of $I_w$. Yellow and white ($\odot|\otimes$) symbols indicate the direction of the ferroelectric polarization for $BiFeO_3$ and magnetization for PtCo, respectively. The blue arrows indicate the direction of the in-plane magnetic field $H_x$, which determines the polarity of magnetization switching from $+M_z$ to $-M_z$ (upper panel) or from $-M_z$ to $+M_z$ (lower panel). The amplitude of $I_w$ is denoted on the left side of each image.

Finally, we discuss the advantages of the demonstrated MMST for in-memory computing. In MMST, two coupled non-volatile state variables—ferroelectric polarization and magnetization (with $I_c$ controlled by ferroelectric polarization)—can be written in parallelism and stored within the memory cell. We exploit this unique functionality and demonstrate reconfigurable Boolean logic computing using a single MMST device. In the 3-terminal device configuration (Fig. 4a), the logic output ($OUT_i$) represented by the magnetization $\pm M_z$ is determined by the logic inputs of applied current $I_w$ (IN) and gate voltage $V_G$ (which controls the ferroelectric polarization acting as the synaptic weight W), as well as the initial magnetization state ($OUT_{i-1}$). By leveraging the non-volatile magnetization state ($OUT_{i-1}$) as the computational operand, a full set of 16 Boolean logic functions can be accomplished using a single MMST memory cell without necessitating changes to the circuit topology (see Supplementary Notes and Supplementary Fig. 7). As discussed earlier, two distinct current thresholds ($I_{c1}$ and $I_{c2}$, while $I_{c1}<I_{c2}$) for switching $M_z$ can be established depending on the ferroelectric polarization direction. As a result, an intermediate $I_w$ ($-I_{c2}<I_w<-I_{c1}$ or $I_{c1}<I_w<I_{c2}$) switches the output magnetization state only when the ferroelectric polarization is upward (W=1). A small $I_w$ ($-I_{c1}<I_w<I_{c1}$) maintains the initial magnetization state, resulting in $OUT_i=OUT_{i-1}$, in which the initial magnetization can be set irrespective of the polarization by a large $I_w$ ($I_w>I_{c2}$ or $I_w<-I_{c2}$). Consequently, complete logic functions can be implemented and reconfigured by defining IN and supplementary step to set the initial state $OUT_{i-1}$. As examples, we present the truth table showing settings of IN and $OUT_{i-1}$ for MMST to function as AND and XNOR logic gates in Fig. 4b, which are the common logic functions required for constructing convolutional neural network[8]. Building upon this configuration, we experimentally demonstrate the operations of a reconfigurable AND (Fig. 4c) and XNOR gate (Fig. 4d) using the SrRuO$_3$/BiFeO$_3$/PtCo device. The output magnetization is monitored using MOKE.

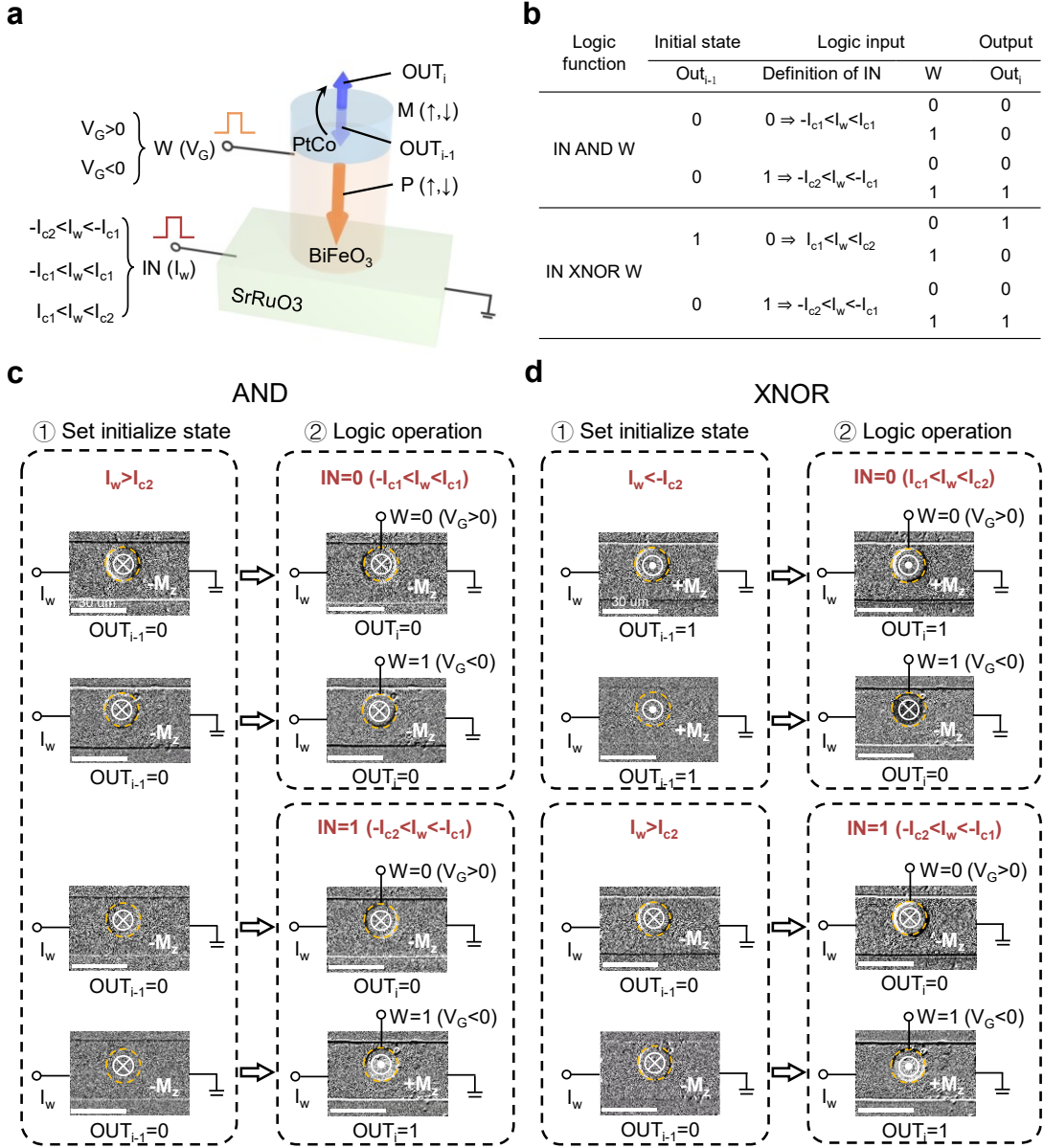

**Figure 4. Reconfigurable logic operations of the MMST device. a**, Schematic illustration of the proposed MMST logic with a single memory cell positioned on a spin-current channel. The logic inputs are the applied current pulse $I_w$ (IN) and the gate voltage pulse $V_G$ (W=0 for $V_G$>0 and W=1 for $V_G$<0). The logic output is represented by the direction of $M_z$ (OUT$_i$=0 for -$M_z$ and OUT$_i$=1 for +$M_z$). Different logic functions can be realized by setting the initial magnetization state (OUT$_{i-1}$) and configuring different amplitudes and polarities of $I_w$ (-$I_{c2}$<$I_w$<-$I_{c1}$, -$I_{c1}$<$I_w$<$I_{c1}$ and $I_{c1}$<$I_w$<$I_{c2}$, where $I_{c1}$ and $I_{c2}$ are threshold currents for switching $M_z$ after applying $V_G$<0 and $V_G$>0,

respectively). **b**, Truth table for the reconfigurable AND and XNOR logic gates. **c**, MOKE images illustrating the AND logic operations in 2 steps. Left, MOKE images for $OUT_{i-1}=0$, set by $I_w>I_{c2}$. Right, MOKE images demonstrate $OUT_i$ with logic inputs of 'IN=0 ($-I_{c1}<I_w<I_{c1}$), W=0 ($V_G>0$)', 'IN=0, W=1 ($V_G<0$)', 'IN=1 ($-I_{c2}<I_w<-I_{c1}$), W=0', and 'IN=1, W=1', respectively. **d**, MOKE images illustrating the XNOR logic operations in 2 steps. Left, MOKE images for $OUT_{i-1}=1$ set by $I_w<-I_{c2}$, and $OUT_{i-1}=0$ set by $I_w>I_{c2}$. Right, MOKE images demonstrate $OUT_i$ with logic inputs of 'IN=0 ($I_{c1}<I_w<I_{c2}$), W=0 ($V_G>0$)', 'IN=0, W=1 ($V_G<0$)', 'IN=1 ($-I_{c2}<I_w<-I_{c1}$), W=0', and 'IN=1, W=1', respectively. The bright and dark contrast in the device correspond to $+M_z$ and $-M_z$, respectively. The logic operations are performed with an in-plane magnetic field.

Thus far, we presented proof-of-concept experiment for reconfigurable logic computing using multiferroic magnons. Remaining reliability issue can be mitigated by mono-domain $BiFeO_3$ that will reduce the leakage current[52]. Supplementary Fig. 8 shows the reversible control of MST by $V_G$ via the modulation of non-collinear antiferromagnetic structure in a monodomain sample[52], rather than the change of ferroelectric domain structures (see PFM in Supplementary Fig. 8), suggesting another possible contribution to the voltage-controlled MST. We anticipate further improvement of the reliability and tunability of the MMST device by domain structures engineering, chemical doping and fabrication process optimizing.

For in-memory-computing applications, the utilization of magnetic tunnel junctions (MTJs) could serve as an electrical read-out mechanism for the output magnetization. The MMST-MTJs can then be incorporated in a crossbar array architecture that uses MTJ resistance summation for high throughput multiply–accumulate (MAC) operations (see Supplementary Notes and Supplementary Fig. 9), a fundamental process in AI. The inherent non-volatility of the ferroelectric logic input and the magnetic logic output allows for the storage of both synaptic weights and intermediate MAC results in a non-volatile manner. This approach significantly minimizes memory area overhead and power consumption by reducing the necessity for intermediate calculation parameter copy and negates the requirement for data reloading after power-off. These features highlight the potential of multiferroic magnonics for low-power neuromorphic computing.

**References**

1. Sebastian, A., Le Gallo, M., Khaddam-Aljameh, R. & Eleftheriou, E. Memory devices and applications for in-memory computing. *Nat. Nanotech.* **15**, 529-544 (2020).

2. Ielmini, D. & Wong, H. S. P. In-memory computing with resistive switching devices. *Nat. Electron.* **1**, 333-343 (2018).

3. Wang, Z. *et al.* Resistive switching materials for information processing. *Nat. Rev. Mater.* **5**, 173-195 (2020).

4. Lanza, M. *et al.* Memristive technologies for data storage, computation, encryption, and radio-frequency communication. *Science* **376**, eabj9979 (2022).

5. Demokritov, S. O. *et al.* Bose-Einstein condensation of quasi-equilibrium magnons at room temperature under pumping. *Nature* **443**, 430-433 (2006).

6. Chumak, A. V., Vasyuchka, V. I., Serga, A. A. & Hillebrands, B. Magnon spintronics. *Nat. Phys.* **11**, 453-461 (2015).

7. Baek, S. C. *et al.* Complementary logic operation based on electric-field controlled spin-orbit torques. *Nat. Electron.* **1**, 398-403 (2018).

8. Jung, S. *et al.* A crossbar array of magnetoresistive memory devices for in-memory computing. *Nature* **601**, 211-216 (2022).

9. Raymenants, E. *et al.* Nanoscale domain wall devices with magnetic tunnel junction read and write. *Nat. Electron.* **4**, 392-398 (2021).

10. Luo, Z. *et al.* Current-driven magnetic domain-wall logic. *Nature* **579**, 214-218 (2020).

11. Manipatruni, S. *et al.* Scalable energy-efficient magnetoelectric spin-orbit logic. *Nature* **565**, 35-42 (2018).

12. Weinstein, A. J. *et al.* Universal logic with encoded spin qubits in silicon. *Nature* **615**, 817-822 (2023).

13. Cornelissen, L. J., Liu, J., Duine, R. A., Youssef, J. B. & van Wees, B. J. Long-distance transport of magnon spin information in a magnetic insulator at room temperature. *Nat. Phys.* **11**, 1022-1026 (2015).

14. Lebrun, R. *et al.* Tunable long-distance spin transport in a crystalline antiferromagnetic iron oxide. *Nature* **561**, 222-225 (2018).

15  Han, J. *et al.* Birefringence-like spin transport via linearly polarized antiferromagnetic magnons. *Nat. Nanotech.* **15**, 563-568 (2020).

16  Guo, C. Y. *et al.* A nonlocal spin Hall magnetoresistance in a platinum layer deposited on a magnon junction. *Nat. Electron.* **3**, 304-308 (2020).

17  Wang, Y. *et al.* Magnetization switching by magnon-mediated spin torque through an antiferromagnetic insulator. *Science* **366**, 1125-1128 (2019).

18  Zhang, S. S. & Zhang, S. Magnon mediated electric current drag across a ferromagnetic insulator layer. *Phys. Rev. Lett.* **109**, 096603 (2012).

19  Wang, H., Du, C., Hammel, P. C. & Yang, F. Antiferromagnonic spin transport from $Y_3Fe_5O_{12}$ into NiO. *Phys. Rev. Lett.* **113**, 097202 (2014).

20  Ross, A. *et al.* Propagation length of antiferromagnetic magnons governed by domain configurations. *Nano Lett.* **20**, 306-313 (2020).

21  Wei, X. Y. *et al.* Giant magnon spin conductivity in ultrathin yttrium iron garnet films. *Nat. Mater.* **21**, 1352-1356 (2022).

22  Han, J., Zhang, P., Hou, J. T., Siddiqui, S. A. & Liu, L. Mutual control of coherent spin waves and magnetic domain walls in a magnonic device. *Science* **366**, 1121-1125 (2019).

23  Cornelissen, L. J., Liu, J., van Wees, B. J. & Duine, R. A. Spin-current-controlled modulation of the magnon spin conductance in a three-terminal magnon transistor. *Phys. Rev. Lett.* **120**, 097702 (2018).

24  Du, C. *et al.* Control and local measurement of the spin chemical potential in a magnetic insulator. *Science* **357**, 195-198 (2017).

25  Qi, S. *et al.* Giant electrically tunable magnon transport anisotropy in a van der Waals antiferromagnetic insulator. *Nat. Commun.* **14**, 2526 (2023).

26  Li, R. *et al.* A puzzling insensitivity of magnon spin diffusion to the presence of 180-degree domain walls. *Nat. Commun.* **14**, 2393 (2023).

27  Fiebig, M., Lottermoser, T., Meier, D. & Trassin, M. The evolution of multiferroics. *Nat. Rev. Mater.* **1**, 16046 (2016).

28  Spaldin, N. A. & Ramesh, R. Advances in magnetoelectric multiferroics. *Nat. Mater.* **18**, 203-212 (2019).

29   Dong, S., Liu, J.-M., Cheong, S.-W. & Ren, Z. Multiferroic materials and magnetoelectric physics: symmetry, entanglement, excitation, and topology. *Adv. Phys.* **64**, 519-626 (2015).

30   Heron, J. T. *et al.* Deterministic switching of ferromagnetism at room temperature using an electric field. *Nature* **516**, 370-373 (2014).

31   Baek, S. H. *et al.* Ferroelastic switching for nanoscale non-volatile magnetoelectric devices. *Nat. Mater.* **9**, 309-314 (2010).

32   de Sousa, R. & Moore, J. E. Electrical control of magnon propagation in multiferroic $BiFeO_3$ films. *Appl. Phys. Lett.* **92**, 022514 (2008).

33   Rovillain, P. *et al.* Electric-field control of spin waves at room temperature in multiferroic $BiFeO_3$. *Nat. Mater.* **9**, 975-979 (2010).

34   Parsonnet, E. *et al.* Nonvolatile electric field control of thermal magnons in the absence of an applied magnetic field. *Phys. Rev. Lett.* **129**, 087601 (2022).

35   Sinova, J., Valenzuela, S. O., Wunderlich, J., Back, C. H. & Jungwirth, T. Spin Hall effects. *Rev. Mod. Phys.* **87**, 1213-1260 (2015).

36   Manchon, A. *et al.* Current-induced spin-orbit torques in ferromagnetic and antiferromagnetic systems. *Rev. Mod. Phys.* **91**, 035004 (2019).

37   Rezende, S. M., Rodríguez-Suárez, R. L. & Azevedo, A. Diffusive magnonic spin transport in antiferromagnetic insulators. *Phys. Rev. B* **93**, 054412 (2016).

38   Bender, S. A., Skarsvåg, H., Brataas, A. & Duine, R. A. Enhanced Spin Conductance of a Thin-Film Insulating Antiferromagnet. *Phys. Rev. Lett.* **119**, 056804 (2017).

39   Zheng, D. *et al.* High-efficiency magnon-mediated magnetization switching in all-oxide heterostructures with perpendicular magnetic anisotropy. *Adv. Mater.* **34**, 2203038 (2022).

40   Gross, I. *et al.* Real-space imaging of non-collinear antiferromagnetic order with a single-spin magnetometer. *Nature* **549**, 252-256 (2017).

41   Zhou, J. *et al.* Modulation of spin-orbit torque from $SrRuO_3$ by epitaxial-strain-induced octahedral rotation. *Adv. Mater.* **33**, 2007114 (2021).

42   Ou, Y. *et al.* Exceptionally high, strongly temperature dependent, spin Hall conductivity of $SrRuO_3$. *Nano Lett.* **19**, 3663-3670 (2019).

43   Chu, Y. H. *et al.* Nanoscale domain control in multiferroic $BiFeO_3$ thin films. *Adv. Mater.* **18**,

2307-2311 (2006).

44  Liu, L. *et al.* Spin-torque switching with the giant spin Hall effect of tantalum. *Science* **336**, 555-558 (2012).

45  Garello, K. *et al.* Symmetry and magnitude of spin-orbit torques in ferromagnetic heterostructures. *Nat. Nanotech.* **8**, 587-593 (2013).

46  Baumgartner, M. *et al.* Spatially and time-resolved magnetization dynamics driven by spin-orbit torques. *Nat. Nanotech.* **12**, 980-986 (2017).

47  Liu, L. *et al.* Current-induced self-switching of perpendicular magnetization in CoPt single layer. *Nat. Commun.* **13**, 3539 (2022).

48  Zhu, L., Ralph, D. C. & Buhrman, R. A. Maximizing spin-orbit torque generated by the spin Hall effect of Pt. *Appl. Phys. Rev.* **8**, 031308 (2021).

49  Lee, K.-S., Lee, S.-W., Min, B.-C. & Lee, K.-J. Threshold current for switching of a perpendicular magnetic layer induced by spin Hall effect. *Appl. Phys. Lett.* **102**, 112410 (2013).

50  Hayashi, M., Kim, J., Yamanouchi, M. & Ohno, H. Quantitative characterization of the spin-orbit torque using harmonic Hall voltage measurements. *Phys. Rev. B* **89**, 144425 (2014).

51  Tang, A. *et al.* Implementing complex oxides for efficient room-temperature spin–orbit torque switching. *Adv. Electron. Mater.* **8**, 2200514 (2022).

52  Saenrang, W. *et al.* Deterministic and robust room-temperature exchange coupling in monodomain multiferroic BiFeO$_3$ heterostructures. *Nat. Commun.* **8**, 1583 (2017).

## Methods

### Sample growth and characterization

Epitaxial SrRuO$_3$/BiFeO$_3$ heterostructures were deposited on (110)$_o$-oriented DyScO$_3$ substrates using pulsed laser deposition (PLD) with a 248 nm KrF excimer laser. The SrRuO$_3$ growth was conducted at a substrate temperature of 670 °C and an oxygen partial pressure of 110 mTorr. The BiFeO$_3$ growth was conducted at a substrate temperature of 700 °C and an oxygen partial pressure of 150 mTorr. The laser fluence at target surfaces was ~1.5 J/cm$^2$ and the pulse repetition was 5-7 Hz. Subsequently, the samples were cooled to room temperature in an oxygen rich atmosphere and transferred to a magnetron sputtering chamber with a background vacuum of $1\times10^{-8}$ Torr for the deposition of ferromagnetic metals. The ferromagnetic multilayer PtCo or NiFe was sputter deposited at an Ar pressure of 3 mTorr. We measured the thickness of films by using x-ray reflectivity.

**Device fabrication**

The samples were patterned by using photolithography followed by Ar ion beam milling for MST measurements. Then electrodes of 100 nm Pt/5 nm Ti were deposited and defined by the lift-off process. Devices for anomalous Hall resistance (shown in Fig. 2) and SHHV (shown in Supplementary Fig. 5) measurements were patterned into 16 μm wide and 80 μm long Hall bars. Devices for ST-FMR measurements (shown in Supplementary Fig. 4) were patterned into 10 μm wide and 50 μm long microstrips with ground-signal-ground electrodes. In devices for voltage-controlled MST with stripe-domain $BiFeO_3$ (shown in Fig. 3 and Fig. 4), the $SrRuO_3$ was patterned into 30 μm wide and 150 μm long microstrips and the $BiFeO_3$/PtCo were patterned into pillars with a diameter of 20-25 μm. In devices for voltage-controlled MST with mono-domain $BiFeO_3$ (shown in Supplementary Fig. 8), the $SrRuO_3$ were patterned into 50 μm wide and 250 μm long microstrips and $BiFeO_3$/PtCo were patterned into pillars with a diameter of 30 μm.

**STEM characterization**

For cross-sectional microscopy, sample was prepared by using focused ion beam (FIB) milling. Cross-sectional lamellas were thinned down to 60 nm thick at an accelerating voltage of 30 kV with a decreasing current from the maximum 2.5 nA, followed by fine polish at an accelerating voltage of 2 kV with a small current of 40 pA. The atomic scale HAADF-STEM images of $SrRuO_3$/$BiFeO_3$/PtCo tri-layer were performed by Cs-corrected JEM ARM200CF microscope operated at 200 kV using a high-angle annular detector for Z-contrast imaging with a collection angle of 90-370 mrad.

**MST-induced switching measurements**

In measurements of MST-induced switching with Hall-bars (shown in Fig. 3), pulsed currents (pulse duration of 1 ms) with various amplitudes were applied to the current channel under an external magnetic field along the current axis. After each pulse, the Hall resistance was measured with a small dc current of 0.5 mA. In measurements of MST-induced switching with the magnetization measured by a polar MOKE microscopy (shown in Fig. 4), pulsed currents (pulse duration of 1 ms) with various amplitudes were applied to the shared spin-current channel under an external magnetic field. The gate voltage pulses $V_G$ were applied to the $BiFeO_3$/PtCo cell by using the shared spin-current channel as the ground and the PtCo as the top electrode. In all measurements, the initial magnetization of PtCo was first set by an out-of-plane external magnetic field. All measurements were performed at room temperature.

**Acknowledgments**

**Funding:** This work was supported by the National Natural Science Foundation (grant numbers 52161135103 and 52250418), Tsinghua University Initiative Scientific Research Program, and the National Key R&D Program of China (grant number 2021YFA0716500). Work at University of Wisconsin-Madison was supported by the funded by Vannevar Bush Faculty Fellowship (ONR N00014-20-1-2844), the Gordon and Betty Moore Foundation's EPiQS Initiative and grant GBMF9065 to C.B.E.. Work at Chinese Academy of Sciences was supported by the National Natural Science Foundation (grant numbers 52250402 and 52025025).

**Author contributions:** T.N., and D.Y. conceived the research. T.N., H.W., C.B.E., D.C.R., D.Y., Y.H.L., J.T., P.Y., J.M., L.G. and W.S. supervised the experiments. Y.W., Y.C., P.P., Y.L. and J.M. performed the sample growth. Y.C., Y.L., D.J., Y.Z. and H.C. performed the device fabrication. Y.C., Y.L., D.J., Y.T. and H.C. performed device measurements and analysis. C.X. performed the circuit model analysis. L.G. and Q.Z. performed the high-resolution STEM experiments. B.L. performed the theoretical calculations. T.N., D.Y., Y.L. and Y.C. wrote the manuscript. All authors discussed the results and commented on the manuscript. T.N., D.Y. and Y.H.L. directed the research.

**Competing interests:** All other authors declare they have no competing interests.

**Data and materials availability:** All data needed to evaluate the conclusions in the paper are present in the paper and/or the Supplementary Materials. Additional data related to this paper may be requested from the authors.

Supplementary materials for

# Multiferroic Magnon Spin-Torque Based Reconfigurable Logic-In-Memory

Yahong Chai[*], Yuhan Liang[*], Cancheng Xiao[*], Yue Wang, Bo Li, Dingsong Jiang, Pratap Pal, Yongjian Tang, Hetian Chen, Yuejie Zhang, Witold Skowroński, Qinghua Zhang, Lin Gu, Jing Ma, Pu Yu, Jianshi Tang, Yuan-Hua Lin[†], Di Yi[†], Daniel C. Ralph, Chang-Beom Eom, Huaqiang Wu, Tianxiang Nan[†]

[†] Corresponding author. Email: linyh@tsinghua.edu.cn, diyi@mail.tsinghua.edu.cn, nantianxiang@mail.tsinghua.edu.cn

# Table of Contents

## Supplementary Notes

**Model for voltage-controlled MST in BiFeO$_3$ with two-variant stripe domain**

We estimate the magnon transport in two-variant stripe domains with spin cycloid, in which spin cycloid propagates along $\vec{k}$ orthogonal to ferroelectric polarization $\vec{P}$ in BiFeO$_3$ (see Supplementary Fig. 1a). For $\vec{k}$ in the x-y plane, the Néel order can be written as:

$$\vec{L}(\mathbf{r}) = \vec{e}_p \cos(\vec{k} \cdot \vec{r}) + \vec{e}_k \sin(\vec{k} \cdot \vec{r}) \tag{1}$$

where $\vec{e}_p$ and $\vec{e}_k$ are the unit vectors of $\vec{P}$ and $\vec{k}$, and $|\mathbf{k}| = 2\pi/\lambda$ with $\lambda$ being the cycloid characteristic wavelength. For domain 1, $\vec{e}_p^1 = \vec{x}\cos\theta_p + \vec{z}\sin\theta_p$ and $\vec{e}_k^1 = \vec{y}$, corresponding to $\vec{k}$ along $\vec{y}$. For domain 2, $\vec{e}_p^2 = -\vec{y}\cos\theta_p + \vec{z}\sin\theta_p$ and $\vec{e}_k^2 = \vec{x}$, corresponding to $\vec{k}$ along $\vec{x}$ (see Supplementary Fig. 1b). Here, $\theta_p = \arctan(1/\sqrt{2})$ and $\vec{x}, \vec{y}$ and $\vec{z}$ are unit vectors. Since our sample is a thin BiFeO$_3$ film, we assume the Néel order distributes uniformly along $\vec{z}$-direction. The averaged magnon current density propagates through the sample is given by[1-3]:

$$\vec{j}^m = \frac{G_m}{\mathcal{A}} \int dx dy \, (\vec{\sigma} \cdot \vec{L}) \vec{L} \tag{2}$$

where $G_m$ is magnon conductance per unit area, $\mathcal{A}$ is the area of x-y plane, and $\vec{\sigma} = \vec{y}$ is the spin polarization direction. The x-y plane is much greater than the cycloid period, i.e. $\mathcal{A} \gg \lambda^2$, thus Eq. (2) can be calculated in one period. By substituting the expression of Néel order of Equation (1) into Equation (2), we obtain the averaged magnon current density in two types of domains:

$$\vec{j}_1^m = \frac{G_m}{2} \vec{e}_k^1 \tag{3}$$

$$\vec{j}_2^m = \frac{G_m}{2} \cos\theta_p \vec{e}_p^2 \tag{4}$$

As the magnon current is injected into ferromagnetic layer, the MST density is given by[4]:

$$\vec{\tau}_m = \vec{M} \times (\vec{j}^m \times \vec{M}) \tag{5}$$

For PtCo, the magnetization $\vec{M}$ is along $\vec{z}$-direction, thus we obtain the magnon torque density as:

$$\tau_m^1 = G_m/2 \qquad (6)$$

$$\tau_m^2 = \frac{G_m}{2} cos^2\theta_p \qquad (7)$$

From the expression of Equation (6) and Equation (7) above, we find that for injected spin polarization $\vec{\sigma}$ along $\vec{y}$, the magnon transport is more efficient through domain 1, leading to a larger MST on magnetization. Thus, the domain population variation after ferroelectric polarization switching can lead to modulation of MST through BiFeO$_3$. Before applying V$_G$, the polarization P$_z$ is downward and two-variant strip domains with orthogonal $\vec{k}$ have equal population (see the left of Supplementary Fig. 1c). After applying V$_G$ to switch the ferroelectric polarization (see the right of Supplementary Fig. 1c), ferroelectric domain wall moves[5], leading to an unequal population of two domains.

**Structure characterization of the SrRuO$_3$/BiFeO$_3$/PtCo and ferroelectricity of BiFeO$_3$ thin film**

We first confirmed the high crystal quality of epitaxial SrRuO$_3$/BiFeO$_3$ heterostructure using X-ray diffraction, showing clear (200)$_{pc}$ Bragg peaks of BiFeO$_3$ and SrRuO$_3$ layers (see Supplementary Fig. 2a). The as-grown SrRuO$_3$/BiFeO$_3$ heterostructure shows ferroelectric two-variant stripe domain, revealed by in-plane PFM phase image (see Supplementary Fig. 2b). To investigate the polarization switching behavior, bias voltages of -5 V was applied to BiFeO$_3$ layer through the PFM tip with the SrRuO$_3$ layer grounded, as shown in Supplementary Fig. 2c. The out-of-plane component of ferroelectric polarization is switched from initially downward to upward (the right of Supplementary Fig. 2c), while the in-plane PFM shows the change of domain population after applying bias voltage (the left of Supplementary Fig. 2c). Electron energy loss spectroscopy (EELS) mapping of elements distribution in an 11 nm SrRuO$_3$/50 nm BiFeO$_3$/PtCo sample is shown in Supplementary Fig. 2d, identifying the sharpness of SrRuO$_3$/BiFeO$_3$ and BiFeO$_3$/PtCo interfaces.

**Additional experimental results of MST-induced switching in SrRuO$_3$/BiFeO$_3$/PtCo and the current-induced self-switching in PtCo**

The structure of the 11 nm SrRuO$_3$/120 nm BiFeO$_3$/PtCo sample for MST-induced switching is shown in Supplementary Fig. 3a. The additional results of MST-induced switching of magnetization under various in-plane magnetic field H$_x$ for Fig. 2 are displayed in Supplementary Fig. 3b. The structure of the control sample (PtCo on Si substrate) is shown in Supplementary Fig. 3c, and corresponding anomalous Hall resistance loop as sweeping out-of-plane magnetic field H$_z$ is shown in Supplementary Fig. 3d. The current-induced switching results under H$_x$ = ± 10 mT for control sample is shown in Supplementary Fig. 3e, and no self-switching effect[6] is observed within the same applied current range of that in SrRuO$_3$/BiFeO$_3$/PtCo sample (Supplementary Fig. 3b). Since all current flows into the PtCo layer, the current density in PtCo layer is much larger than that in the SrRuO$_3$/ BiFeO$_3$/PtCo sample. The Hall bar size of PtCo sample is same with that of the SrRuO$_3$/ BiFeO$_3$/PtCo sample.

**Experimental results of ST-FMR measurements**

The ST-FMR measurements were carried out at room temperature. During ST-FMR measurements, a microwave current I$_{rf}$ at a fixed frequency was applied with the in-plane magnetic field swept from 0 to 0.25 T to drive the resonance of the ferromagnetic layer NiFe. The representative optical image of the device for the ST-FMR measurement is shown in Supplementary Fig. 4a. The amplitude of the microwave current is modulated at a low frequency (1.713 kHz), and the mixing voltage $V_{mix}$ is detected through a lock-in amplifier. The resonance spectra were obtained at a fixed frequency with sweeping an in-plane external magnetic field H$_{ext}$. Typical ST-FMR results of various microwave frequencies for 11 nm SrRuO$_3$/120 nm BiFeO$_3$/5 nm NiFe are shown in Supplementary Fig. 4b. The device resistance as a function of magnetic field angle φ due to the anisotropic magnetoresistance (AMR) of NiFe is shown in Supplementary Fig. 4c. The ST-FMR $V_{mix}$ is a combination of I$_{rf}$ and AMR, which can be then written in the form as[7]:

$$V_{mix} = V_S \frac{W^2}{(\mu_0 H_{ext} - \mu_0 H_{FMR})^2 + W^2} + V_A \frac{W(\mu_0 H_{ext} - \mu_0 H_{FMR})}{(\mu_0 H_{ext} - \mu_0 H_{FMR})^2 + W^2} \quad (8)$$

where W is the half-width-at-half-maximum resonance linewidth, and H$_{FMR}$ is the resonance field. $V_S$ and $V_A$ are the symmetric and antisymmetric amplitude of the Lorentzian, and can be expressed as[8]:

$$V_S = -\frac{I_{rf}}{2}\left(\frac{dR}{d\varphi}\right)\frac{1}{\alpha(2\mu_0 H_{FMR}+\mu_0 M_{eff})}\tau_{m,DL} \quad (9)$$

$$V_A = -\frac{I_{rf}}{2}\left(\frac{dR}{d\varphi}\right)\frac{\sqrt{1+M_{eff}/H_{FMR}}}{\alpha(2\mu_0 H_{FMR}+\mu_0 M_{eff})}\tau_{m,FL} \quad (10)$$

where $I_{rf}$ is the microwave current, R($\varphi$) is the device resistance as a function of angle $\varphi$ of in-plane magnetic field based on the AMR, and $\mu_0 M_{eff}$ is the effective magnetization of NiFe and $\alpha$ is the Gilbert damping. Typical ST-FMR spectrum at 7 GHz with $\varphi$=130° is shown in Supplementary Fig. 4d. The symmetric and antisymmetric components can be obtained by fitting the spectrum to symmetric and antisymmetric Lorentzian functions according to Equation (8) to Equation (10). Moreover, the magnetic field angle $\varphi$ dependence of $V_S$ and $V_A$ components both follow $sin2\varphi cos\varphi$ dependence (see Supplementary Fig. 4e), showing the magnon current carries spin polarization along $\vec{y}$.

The magnitude of magnon-torque components can be determined by extracting the symmetric and antisymmetric amplitude from Equation (9) and Equation (10), expressed as[9]:

$$\xi_{m,FL/m,DL} = \tau_{m,FL/m,DL}\, M_s t \frac{l}{I_{rf}R\cos(\varphi)}\left(\frac{2e}{\hbar}\right)\rho \quad (11)$$

where $M_s$ and $t$ are the saturation magnetization and the thickness of NiFe, $l$ is the length of the device bar, $\hbar$ is the reduced Planck's constant, e is the electron charge. The effective magnetization $M_{eff}$ is obtained by measuring the frequency dependent $H_{FMR}$ (see Supplementary Fig. 4f), with a fit to Kittel equation expressed as[8]:

$$2\pi f = \gamma\sqrt{H_{res}(H_{res}+4\pi M_{eff})} \quad (12)$$

where $\mu_0 H_k$ is the in-plane magnetic anisotropy field and $\gamma$ is the gyromagnetic ratio. We find $\xi_{m,DL}$ = 0.012 at room temperature for the 11 nm SrRuO$_3$/120 nm BiFeO$_3$/5 nm NiFe via ST-FMR measurements.

**Experimental results of SHHV measurements**

The SHHV measurements were conducted on the 11 nm SrRuO$_3$/120 nm BiFeO$_3$/5 nm NiFe. By applying an ac current $I_{ac}$, the magnetization can be oscillated by field-like torque $\tau_{m,FL}$ and damping-like torque $\tau_{m,DL}$ around the equilibrium position, resulting in second harmonic hall voltage $V_{2\omega}$, as shown in Supplementary Fig. 5a. The $V_{2\omega}$ is measured while an external magnetic field $H_{ext}$ is rotated within the sample plane, to different angles $\varphi$ with respect to $I_{ac}$. The measured $V_{2\omega}$ is fitted against $\varphi$ using[10,11]:

$$V_{2\omega} = I_{ac}\frac{R_{PHE}\tau_{m,FL}}{\gamma H_{ext}}cos(2\varphi)cos\varphi + I_{ac}[\frac{R_{AHE}\tau_{m,DL}}{2\gamma(\mu_0 M_s + H_{ext})} + R_{ANE}]cos\varphi + I_{ac}R_x sin(2\varphi) \quad (13)$$

where $R_{PHE}$, $R_{AHE}$, and $\gamma$ are the planar Hall resistance, the anomalous Hall resistance, and the gyromagnetic ratio, respectively. The $cos(2\varphi)cos\varphi$ term describes the strength of $\tau_{m,FL}$. The $cos\varphi$ term consists of contributions from both the $\tau_{m,DL}$ and the anomalous Nernst effect ($V_{ANE}$). The $sin2\varphi$ term is proportional to resistance $R_x$, and plays a minor role in the $V_{2\omega}$[11]. The typical $V_{2\omega}$ result is shown in Supplementary Fig. 5b, which can be well fitted using Eq. (13). We note that, the $R_{PHE}$ was obtained by monitoring first harmonic voltage $V_{1\omega}$ as rotating $H_{ext}$ within the sample plane (see Supplementary Fig. 5c) and the $V_{2\omega}$ spectrum with various magnitude of $H_{ext}$ were also detected (see Supplementary Fig. 5d). Then, we plot $I_{ac}\frac{R_{PHE}\tau_{m,FL}}{\gamma H_{ext}}$ against $\frac{1}{H_{ext}}$ to extract the $\tau_{m,FL}$ from the gradient of the linear fit (see Supplementary Fig. 5e). After removing the constants from the amplitude of the $cos\varphi$ term, we plot $I_{ac}\frac{R_{AHE}\tau_{m,DL}}{2\gamma(\mu_0 M_s + H_{ext})}$ against $\frac{1}{(\mu_0 M_s + H_{ext})}$ to extract the $\tau_{m,DL}$ from the slope of the linear fit (see Supplementary Fig. 5f). The efficiency of damping-like MST is estimated using $\xi_{m,DL} = \frac{M_s t_{FM}}{\hbar/2e} \cdot \frac{\tau_{m,DL}}{j_c}$, where $j_c$ is the current density in the SrRuO$_3$ layer (about 1.1×10$^6$ A/cm$^2$.), and $t_{FM}$ is the thickness of the NiFe layer. We find $\xi_{m,DL}$ = 0.027 at room temperature for the 11 nm SrRuO$_3$/120 nm BiFeO$_3$/5 nm NiFe via SHHV measurements.

**Experimental results of polar MOKE measurements for the voltage-controlled magnetic anisotropy effect.**

To exclude voltage-controlled the magnetic anisotropy effect[12], we measured the out-of-plane polar MOKE signals as a function of out-of-plane magnetic field $H_z$ for the 11 nm SrRuO$_3$/100 nm BiFeO$_3$/PtCo sample before and after applying $V_G$ at the same position of the pillar (as shown in Fig. 3). The polar MOKE hysteresis loops are shown in Supplementary Fig. 6, and we observed negligible changes in the hysteresis loop of the cell before and after $V_G$ application, which excludes the voltage-controlled magnetic anisotropy effect for the MST switching.

**Definitions of IN and settings of $OUT_{i-1}$ for performing all 16 Boolean logic functions**

The charge current pulse $I_w$ with different direction and amplitude ($-I_{c2}<I_w<-I_{c1}$, $-I_{c1}<I_w<I_{c1}$ or $I_{c1}<I_w<I_{c2}$), polarized state ("↓" or "↑") and magnetization state ("↓" or "↑") represent IN, W (Weight) and $OUT_i/OUT_{i-1}$, respectively. W can be switched by voltage pulse. $OUT_{i-1}$ is the state variable of ferromagnetic layer which should be initialize "↓" ($I_w>I_{c2}$) or "↑" ($I_w<-I_{c2}$) dependent on the requirements of different logic and IN. The logic operation can be performed with different current pulse $I_w$. Hence, the definitions of IN and initialization of $OUT_{i-1}$ for performing different Boolean logic functions can be ascertained by logical derivation and the lookup of truth table. For the most Boolean logic functions, searching directly in truth table is an easier way to complete the ascertainment. In another way, the logic equation, reflecting the writing behavior of device, can be derivation with simplification with Karnaugh map as:

$$OUT_i = \bar{S} \cdot OUT_{i-1} + \bar{A} \cdot OUT_{i-1} + \bar{W} \cdot OUT_{i-1} + \bar{S} \cdot A \cdot W \quad (14)$$

where the $S$ and $A$ are the symbol ("1" for positive and "0" for negative) and amplitude ("1" for $|I_{c1}|<|I_w|<|I_{c2}|$ and "0" for $0<|I_w|<|I_{c1}|$) of $I_w$, respectively. The combination of $S$ and $A$ has four possible values, where "00" and "10" are equivalent. The definition of IN and initialization of $OUT_{i-1}$ can be clarified by presetting $OUT_{i-1}$ and constructing the objective function (see Supplementary Fig. 7).

**Reversible Voltage control of MST-induced switching in a sample with the mono-domain $BiFeO_3$**

To realize reversible voltage-control MST, we fabricate 15 nm $SrRuO_3$/150 nm $BiFeO_3$/PtCo on (100) $SrTiO_3$ single-crystal substrates with 4° miscut toward $[110]_{pc}$ direction. Due to the epitaxy strain and 4° miscut of $SrTiO_3$ substrate, the $BiFeO_3$ layer shows ferroelectric mono-domain structure, revealed by PFM phase images[12] (Supplementary Fig. 8a and Supplementary Fig. 8b).

The out-of-plane PFM phase signal hysteresis loop is monitored by sweeping out-of-plane applied voltage, indicating the coercive voltage about 10 V to switch the ferroelectric polarization, as shown in Supplementary Fig. 8c. Then, *in-situ* voltage-controlled MST-induced switching of magnetization was conducted. The $I_w$ was applied to spin-current channel after applying $V_G$ to PtCo/BiFeO$_3$ pillar, with probe tip keeping contact with pillar (see Materials and Methods for the device fabrication). An external magnetic field was applied during the MST-induced switching. The pulse width of $I_w$ is 1 ms. We found that $I_{th}$ =12.0 mA (12.5 mA) after applied $V_G$=+20 V (-20 V) for several cycles, as shown in Supplementary Fig. 8d. The corresponding polar MOKE images for MST switching with various $I_w$ after applying $V_G$ are displayed in Supplementary Fig. 8e.

**MMST-MTJs crossbar array and the equivalent circuits**

Built on the MMST device in the main text, we can demonstrate a MMST-MTJs crossbar array for high throughput multiply–accumulate (MAC) operations. The schematic is shown in Supplementary Fig. 9a, which consists of n MMST-MTJ cells stacked on a write bit line for a shared write current. Each cell is controlled by a gate voltage through transistor. When the multiplier and accumulation operations are performed, the results of each basic operation are obtained via the reconfigurable characteristics of the MMST device and stored in the MTJ. Then the same voltage is applied to each source line. In this way, the reading line accumulates the conductance of each MTJ on the same column to obtain the total current, which is converted to the final digital result using an analog-to-digital converter (ADC). The equivalent circuit of a MMST-MTJs column for MAC operation is also shown in Supplementary Fig. 9b. Two non-volatile variable states of ferroelectric polarization and conductance of MTJ are equivalent to two registers in each MMST-MTJ that are utilized to store the synaptic weight and intermediate calculation result, respectively. Different types of logic operations that can be implemented in the MMST device (as shown in the Supplementary Fig. 7) is equivalent to the reconfigurable gate. The intermediate calculation results are accumulated via the ADC which is equivalent to an adder.

**Supplementary Figures**

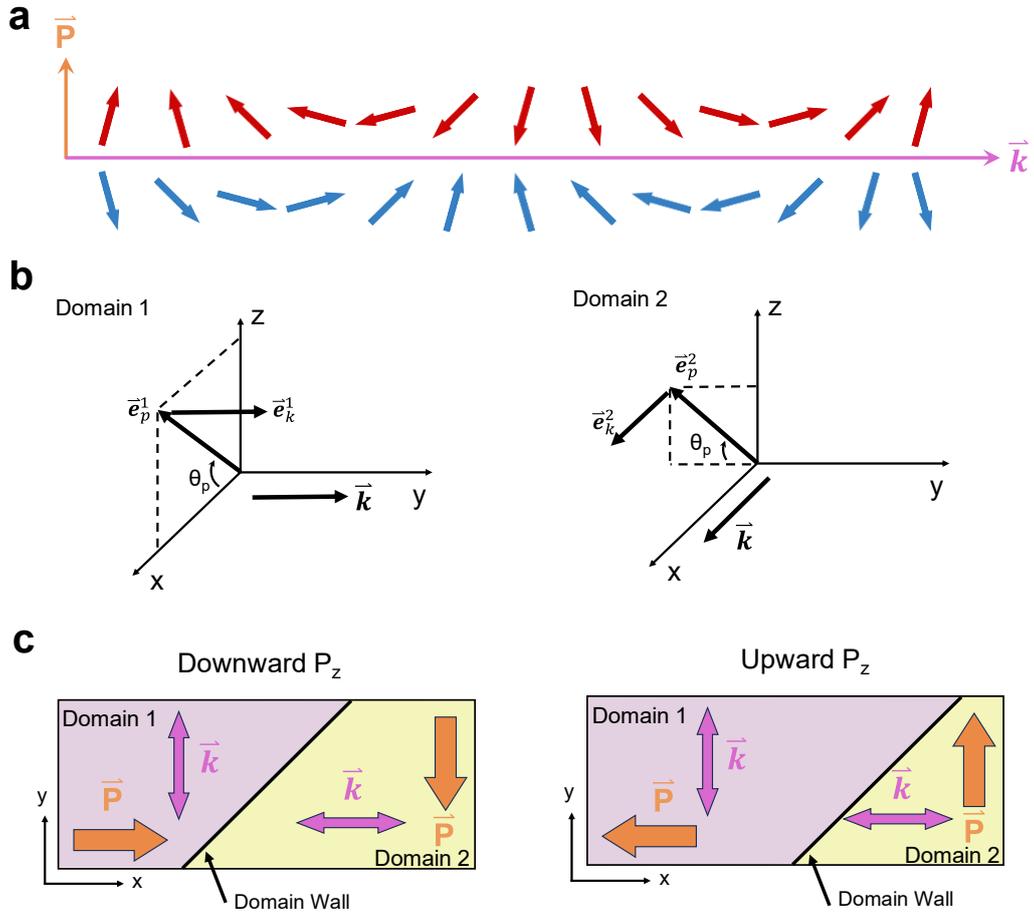

**Supplementary Figure 1. Schematic illustration of the model for voltage-controlled MST in BiFeO$_3$ with two-variant stripe domain structure. a**, Schematic of spin cycloid structure in BiFeO$_3$ on DyScO$_3$. The red and blue arrows represent the two sublattice magnetic moments. **b**, Néel order in domain 1 and domain 2. The cycloid propagation direction $\vec{k}$ of each domain is denoted. **c**, Schematic of ferroelectric domain structure change before and after applying V$_G$. Purple region for domain 1 and yellow region for domain 2.

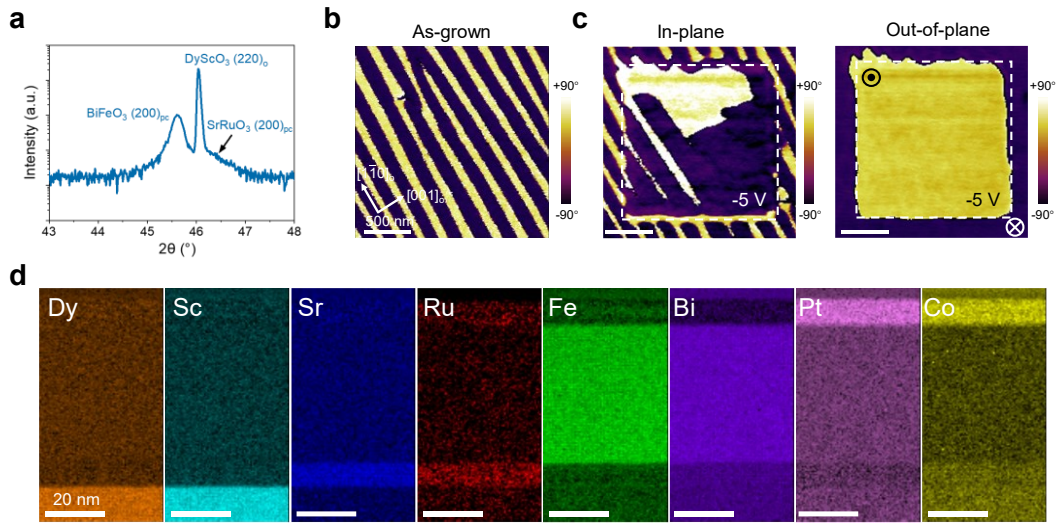

**Supplementary Figure 2. Structure characterization of the SrRuO$_3$/BiFeO$_3$/PtCo and ferroelectricity of BiFeO$_3$ thin film. a**, 2$\theta$-$\omega$ x-ray diffraction scan of SrRuO$_3$/BiFeO$_3$/PtCo heterostructure on (110)$_o$ DyScO$_3$ substrate zoomed along the (220)$_o$ peak of DyScO$_3$. **b**, In-plane PFM phase image of the SrRuO$_3$/BiFeO$_3$ sample taken at the as-grown state, showing two-variant stripe domain structure. **c**, In-plane (left) and out-of-plane (right) PFM phase images of the SrRuO$_3$/BiFeO$_3$ sample taken after the switching written by negative bias voltage of -5 V (area squared by white dashed line). **d**, Electron energy loss spectroscopy (EELS) mapping of elements distribution in an 11 nm SrRuO$_3$/50 nm BiFeO$_3$/PtCo sample.

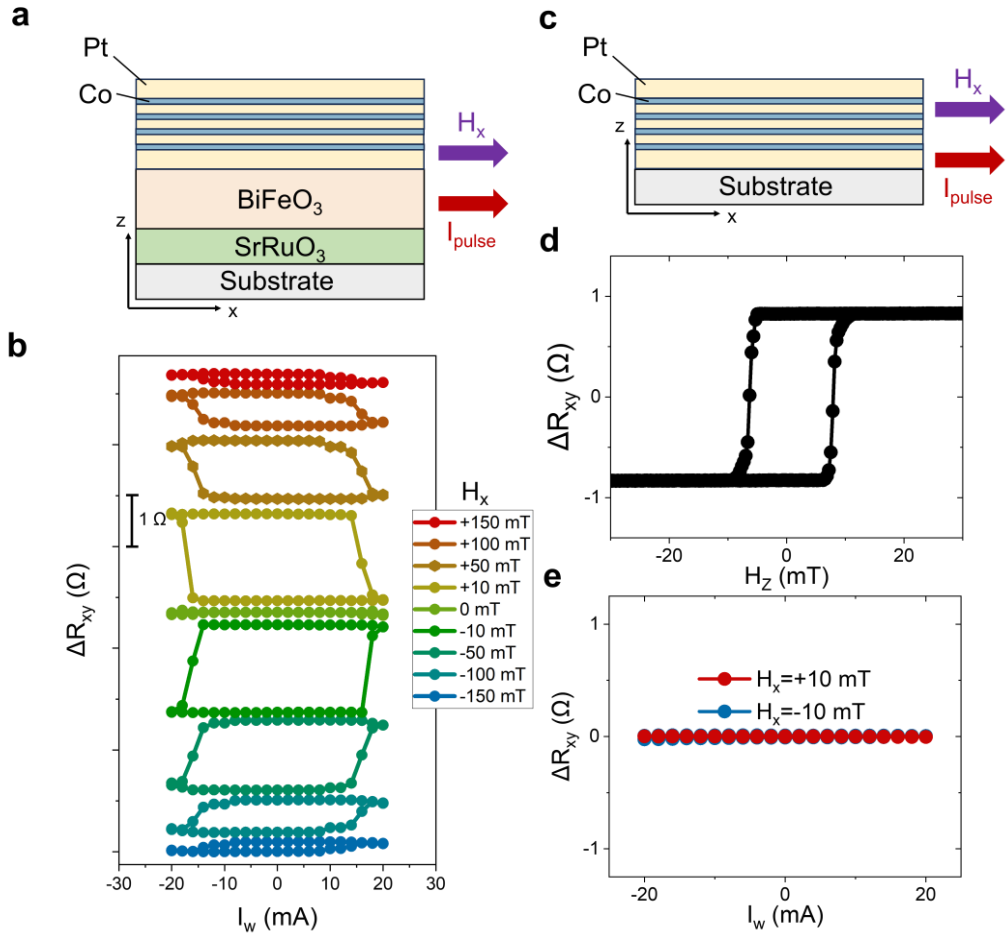

**Supplementary Figure 3. Additional experimental results of MST-induced switching in SrRuO$_3$/BiFeO$_3$/PtCo and the current-induced self-switching in PtCo. a**, Schematic of the layered structure for the 11 nm SrRuO$_3$/120 nm BiFeO$_3$/PtCo. **b**, MST-induced switching of magnetization under various in-plane magnetic field $H_x$ for the 11 nm SrRuO$_3$/120 nm BiFeO$_3$/PtCo. The anomalous Hall resistance $\Delta R_{xy}$-$I_w$ loops are manually shifted for better visualization. **c**, Schematic of the layered structure for the PtCo on a Si substrate. **d**, Anomalous Hall resistance loop for PtCo as sweeping out-of-plane magnetic field $H_z$. **e**, The anomalous Hall resistance $\Delta R_{xy}$-$I_w$ loops for PtCo under $H_x=\pm10$ mT.

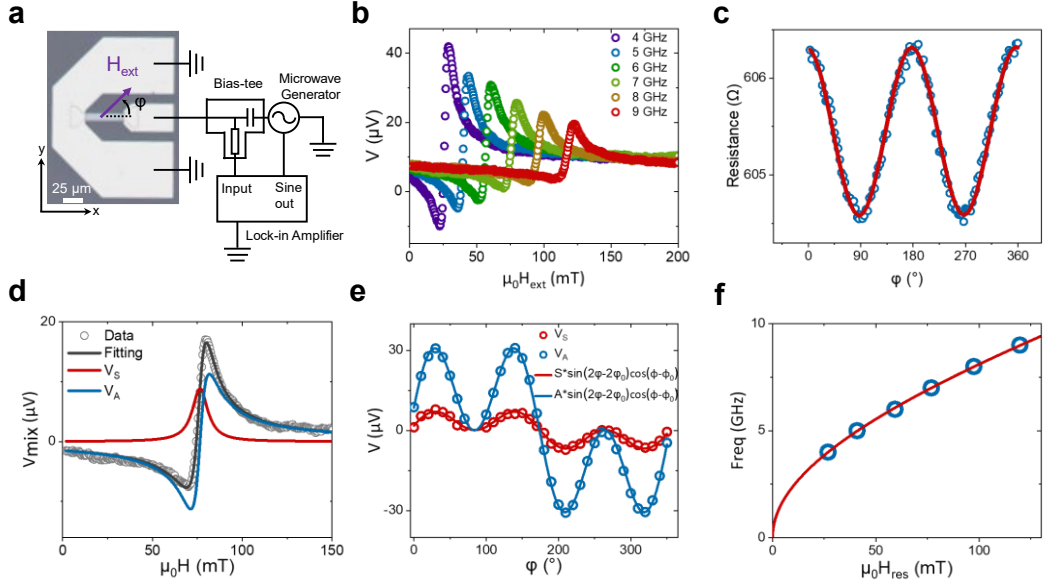

**Supplementary Figure 4. Experimental results of ST-FMR measurements. a**, Optical micrograph and the experimental setup of ST-FMR measurement. **b**, ST-FMR spectra of a 11 nm SrRuO$_3$/120 nm BiFeO$_3$/5 nm NiFe device measured at various frequencies. **c**, Device resistance as a function of magnetic field angle. The blue circles present the experimental data, the red curve represent the $sin2\varphi$ fitting curve. **d**, ST-FMR spectrum fitted to a Lorentzian function. The red and blue lines show the fits of the symmetric V$_S$ and antisymmetric V$_A$ components, respectively. **e**, Symmetric V$_S$ and antisymmetric V$_A$ components as a function of magnetic field angle $\varphi$ at 7 GHz. The data is fitted to $sin2\varphi cos\varphi$. **f**, Resonance frequency as function of the resonance field ($\mu_0 H_{res}$), which is fitted to the Kittel equation.

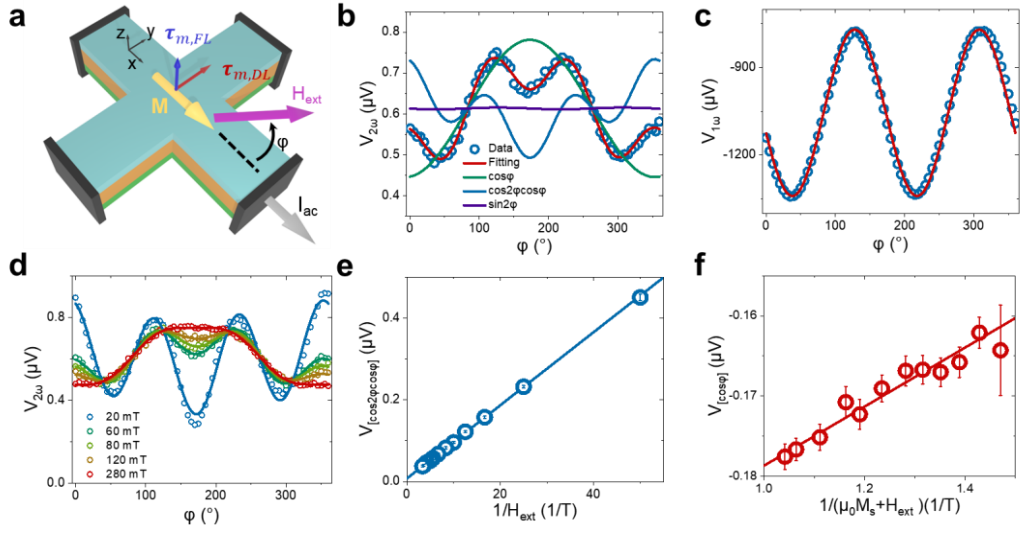

**Supplementary Figure 5. Experimental results of SHHV measurements. a**, Schematic diagram of the experimental setup for SHHV measurement. **b**, $V_{2\omega}$ as a function of the magnetic field angle $\varphi$ with $H_{ext}$ =80 mT for the 11 nm $SrRuO_3$/ 120 nm BFO/5nm NiFe. The circles represent the experimental data, and red curve represents the fit to Equation (13). The green, blue and purple curves represent the fits of $cos\varphi$, $cos2\varphi cos\varphi$ and $sin2\varphi$ components, respectively. **c**, $V_{1\omega}$ as a function of the magnetic field angle $\varphi$ with $H_{ext}$ =80 mT for the same device. The circles present the experimental data, and the red curve represents the fitting curve to $cos2\varphi$. **d**, Second-harmonic Hall signals $V_{2\omega}$ as a function of the magnetic field angle φ with various field amplitudes for the same device. **e**, $cos2\varphi cos\varphi$-dependent $V_{2\omega}$ component plotted as a function of $1/H_{ext}$, where the error bars are due to the uncertainty of the fitting. **f**, $cos\varphi$-dependent $V_{2\omega}$ component plotted as a function of $\frac{1}{(\mu_0 M_s + H_{ext})}$, where the error bars are due to the uncertainty of the fitting.

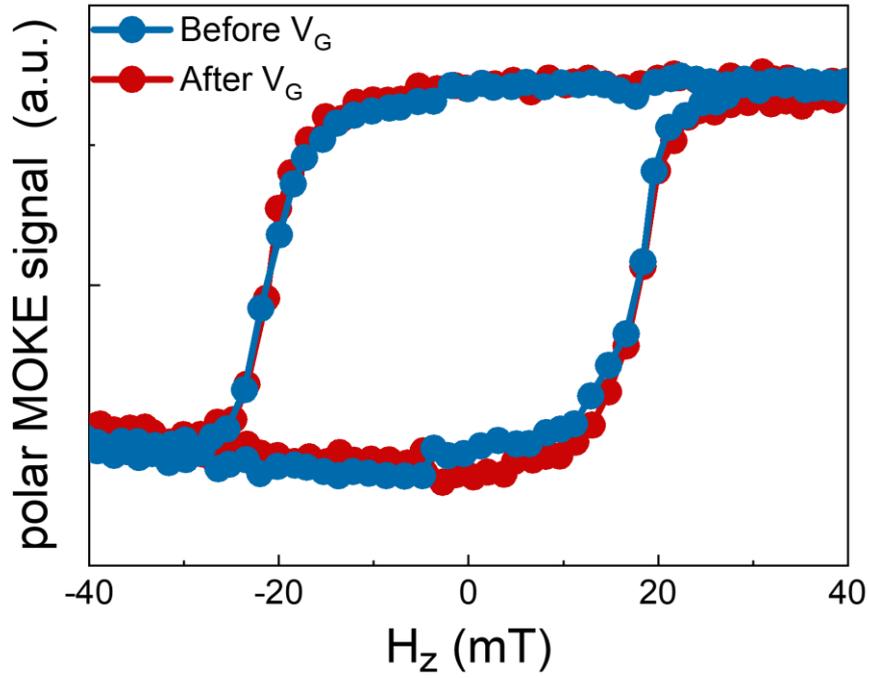

**Supplementary Figure 6. Experimental results of polar MOKE measurements for the voltage-controlled magnetic anisotropy effect.** Polar MOKE intensity obtained from the PtCo cell (as shown in Fig. 3) as a function of out-of-plane magnetic field $H_z$ for the 11 nm $SrRuO_3$/100 nm $BiFeO_3$/PtCo sample before and after applying $V_G$.

|  | Logic function | Initial state ($Out_{i-1}$) | Definition of IN |
|---|---|---|---|
| 1 | TRUE | 1 | 1 ⇒ $-I_{c1}<I_w<I_{c1}$ |
|  |  | 1 | 0 ⇒ $-I_{c1}<I_w<I_{c1}$ |
| 2 | FALSE | 0 | 1 ⇒ $-I_{c1}<I_w<I_{c1}$ |
|  |  | 0 | 0 ⇒ $-I_{c1}<I_w<I_{c1}$ |
| 3 | IN | 1 | 1 ⇒ $-I_{c1}<I_w<I_{c1}$ |
|  |  | 0 | 0 ⇒ $-I_{c1}<I_w<I_{c1}$ |
| 4 | not IN | 0 | 1 ⇒ $-I_{c1}<I_w<I_{c1}$ |
|  |  | 1 | 0 ⇒ $-I_{c1}<I_w<I_{c1}$ |
| 5 | W | 1 | 1 ⇒ $-I_{c1}<I_w<I_{c1}$ |
|  |  | 0 | 0 ⇒ $-I_{c1}<I_w<I_{c1}$ |
| 6 | not W | 0 | 1 ⇒ $-I_{c1}<I_w<I_{c1}$ |
|  |  | 1 | 0 ⇒ $-I_{c1}<I_w<I_{c1}$ |
| 7 | IN AND W | 0 | 1 ⇒ $-I_{c2}<I_w<-I_{c1}$ |
|  |  | 0 | 0 ⇒ $-I_{c1}<I_w<I_{c1}$ |
| 8 | IN NAND W | 1 | 1 ⇒ $I_{c1}<I_w<I_{c2}$ |
|  |  | 1 | 0 ⇒ $-I_{c1}<I_w<I_{c1}$ |
| 9 | IN OR W | 1 | 1 ⇒ $-I_{c1}<I_w<I_{c1}$ |
|  |  | 0 | 0 ⇒ $-I_{c2}<I_w<-I_{c1}$ |
| 10 | IN NOR W | 0 | 1 ⇒ $-I_{c1}<I_w<I_{c1}$ |
|  |  | 1 | 0 ⇒ $I_{c1}<I_w<I_{c2}$ |
| 11 | IN IMP W | 0 | 1 ⇒ $-I_{c2}<I_w<-I_{c1}$ |
|  |  | 1 | 0 ⇒ $-I_{c1}<I_w<I_{c1}$ |
| 12 | IN NIMP W | 1 | 1 ⇒ $I_{c1}<I_w<I_{c2}$ |
|  |  | 0 | 0 ⇒ $-I_{c1}<I_w<I_{c1}$ |
| 13 | IN RIMP W | 1 | 1 ⇒ $-I_{c1}<I_w<I_{c1}$ |
|  |  | 1 | 0 ⇒ $I_{c1}<I_w<I_{c2}$ |
| 14 | IN RNIMP W | 0 | 1 ⇒ $-I_{c1}<I_w<I_{c1}$ |
|  |  | 0 | 0 ⇒ $-I_{c2}<I_w<-I_{c1}$ |
| 15 | IN XOR W | 1 | 1 ⇒ $I_{c1}<I_w<I_{c2}$ |
|  |  | 0 | 0 ⇒ $-I_{c2}<I_w<-I_{c1}$ |
| 16 | IN XNOR W | 0 | 1 ⇒ $-I_{c2}<I_w<-I_{c1}$ |
|  |  | 1 | 0 ⇒ $I_{c1}<I_w<I_{c2}$ |

**Supplementary Figure 7. Definitions of IN and settings of $OUT_{i-1}$ for performing all 16 Boolean logic functions.** Different logic functions can be realized by setting the initial magnetization state ($OUT_{i-1}$) and configuring different amplitudes and polarities of $I_w$ ($-I_{c2}<I_w<-I_{c1}$, $-I_{c1}<I_w<I_{c1}$ and $I_{c1}<I_w<I_{c2}$).

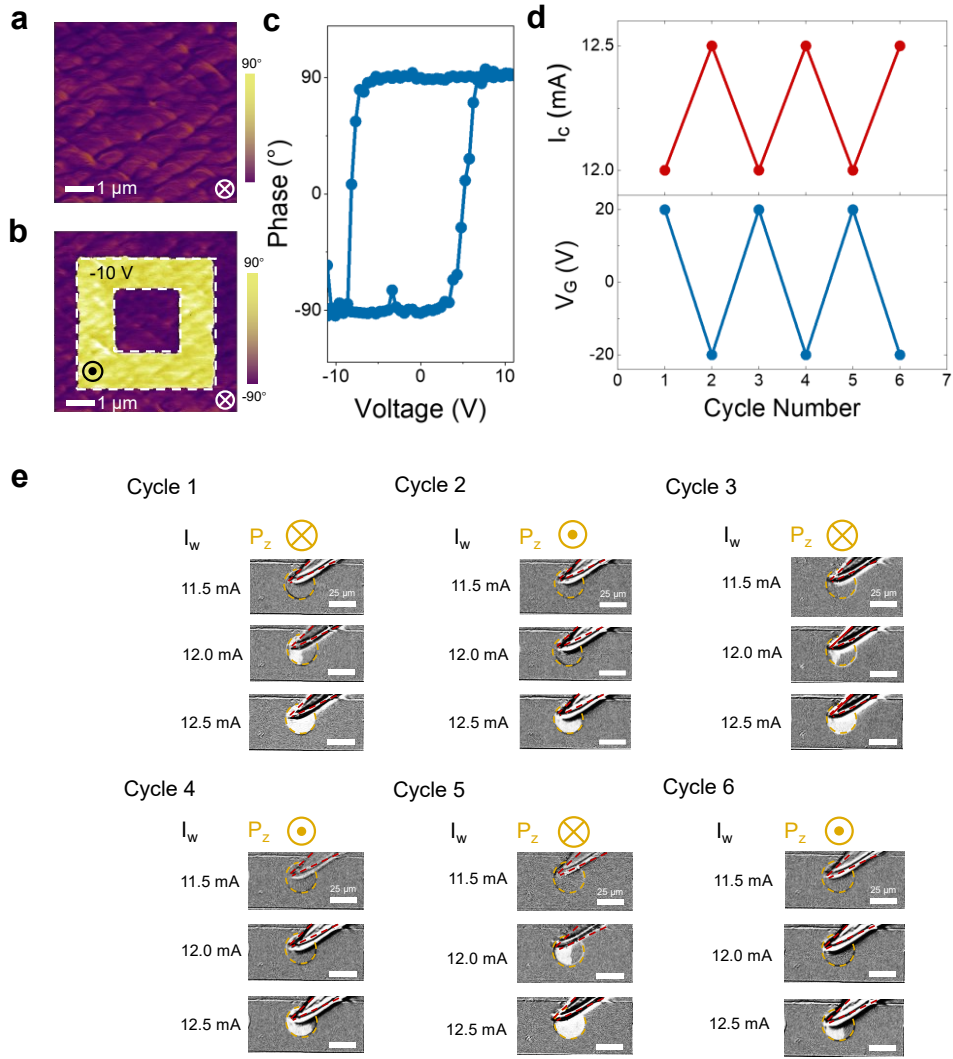

**Supplementary Figure 8. Voltage control of MST-induced switching in a sample with the mono-domain BiFeO₃.** **a-c**, Ferroelectricity of the mono-domain BiFeO₃ in a 15 nm SrRuO₃/150 nm BiFeO₃. **a**, Out-of-plane PFM phase image of the as-grown sample. **b**, Out-of-plane PFM phase image written by bias voltages of -10 V. **c**, Out-of-plane PFM phase signal of PFM as a function of the applied voltage. **d**, Reversible modulation of the threshold current $I_{th}$ for MST-induced switching after applying $V_G$ for several cycles. **e**, Polar MOKE images of MST-induced switching for each cycle shown in **d**. The PtCo cells are denoted by yellow dash circles. Yellow ⊙|⊗ symbols indicate the direction of ferroelectric polarization of BiFeO₃. The red dash lines indicate the probe tip for applying $V_G$.

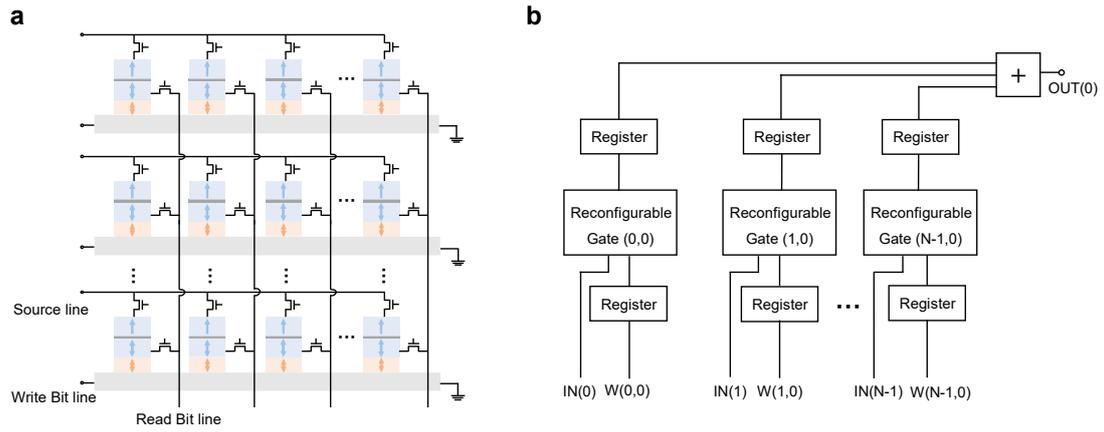

**Supplementary Figure 9. MMST-MTJs crossbar array and the equivalent circuits. a**, Schematic of MMST-MTJs crossbar array. **b**, Equivalent circuit of a MMST-MTJs column for the MAC operation.

**Supplementary References**

1   Qiu, Z. *et al.* Spin colossal magnetoresistance in an antiferromagnetic insulator. *Nat. Mater.* **17**, 577-580 (2018).

2   Rezende, S. M., Rodríguez-Suárez, R. L. & Azevedo, A. Diffusive magnonic spin transport in antiferromagnetic insulators. *Phys. Rev. B* **93**, 054412 (2016).

3   Bender, S. A., Skarsvag, H., Brataas, A. & Duine, R. A. Enhanced Spin Conductance of a Thin-Film Insulating Antiferromagnet. *Phys. Rev. Lett.* **119**, 056804 (2017).

4   Zheng, D. *et al.* High-efficiency magnon-mediated magnetization switching in all-oxide heterostructures with perpendicular magnetic anisotropy. *Adv. Mater.* **34**, 2203038 (2022).

5   Chu, Y. H. *et al.* Nanoscale domain control in multiferroic $BiFeO_3$ thin films. *Adv. Mater.* **18**, 2307-2311 (2006).

6   Liu, L. *et al.* Current-induced self-switching of perpendicular magnetization in CoPt single layer. *Nat. Commun.* **13**, 3539 (2022).

7   Nan, T. *et al.* Comparison of spin-orbit torques and spin pumping across NiFe/Pt and NiFe/Cu/Pt interfaces. *Phys. Rev. B* **91**, 214416 (2015).

8   MacNeill, D. *et al.* Control of spin-orbit torques through crystal symmetry in $WTe_2$/ferromagnet bilayers. *Nat. Phys.* **13**, 300-305 (2017).

9   Nan, T. *et al.* Anisotropic spin-orbit torque generation in epitaxial SrIrO3 by symmetry design. *Proc. Natl. Acad. Sci. U.S.A.* **116**, 16186-16191 (2019).

10  Kang, M. G. *et al.* Electric-field control of field-free spin-orbit torque switching via laterally modulated Rashba effect in Pt/Co/$AlO_x$ structures. *Nat. Commun.* **12**, 7111 (2021).

11  Avci, C. O. *et al.* Interplay of spin-orbit torque and thermoelectric effects in ferromagnet/normal-metal bilayers. *Phys. Rev. B* **90**, 224427 (2014).

12  Saenrang, W. *et al.* Deterministic and robust room-temperature exchange coupling in monodomain multiferroic $BiFeO_3$ heterostructures. *Nat. Commun.* **8**, 1583 (2017).